\documentclass[nofootinbib,twocolumn]{revtex4-2}

\usepackage{graphicx,rotating,amsmath,amsfonts,amssymb}
\usepackage{hyperref}
\usepackage{orcidlink}

\newcommand{\be}{\begin{equation}}
\newcommand{\ee}{\end{equation}}
\newcommand{\benn}{\begin{equation*}}
\newcommand{\eenn}{\end{equation*}}
\newcommand{\bea}{\begin{eqnarray}}
\newcommand{\eea}{\end{eqnarray}}
\newcommand{\beann}{\begin{eqnarray*}}
\newcommand{\eeann}{\end{eqnarray*}}

\newcommand{\perMpccube}{{h^3~{\rm Mpc}}^{-3}}
\newcommand{\hMsun}{h^{-1}{\rm M}_\odot}
\newcommand{\ie}{{\emph i.e.}}
\newcommand{\eg}{{\emph e.g.}}
\newcommand{\dd}{\mathrm{d}}

\begin{document}

\title{Clustering of JWST LRDs. Implications for PBHs and other early-forming seeds}

\author{Gert H\"utsi\,\orcidlink{0000-0002-9322-004X}}
\email{gert.hutsi@kbfi.ee}
\affiliation{Laboratory of High Energy and Computational Physics, National Institute of Chemical Physics and Biophysics, R{\"a}vala pst.~10, 10143, Tallinn, Estonia}


\begin{abstract}
We use the observed abundance and spatial clustering of JWST little red dots (LRDs) over $z = 4$--8 to confront two formation scenarios: the standard picture in which LRDs form in virialized dark-matter halos, and one in which they are seeded by a dynamically subdominant population of primordial black holes (PBHs). The halo formation scenario accounts for both observables naturally and without fine-tuning. In the PBH scenario, matching the observed LRD abundance requires a negligibly small PBH mass fraction, and a power-spectrum analysis shows that such a Poisson isocurvature component predicts a clustering signal far below what is observed. The same conclusion holds for any population of dynamically subdominant early-forming seeds, unless their initial clustering is finely tuned or their activity is tied to massive host halos. We conclude that the spatial clustering of JWST LRDs strongly favours the standard halo formation picture and disfavours PBHs, or similar early-forming seeds, as a randomly activated LRD population; PBHs remain viable as seeds only if their activity is governed by massive host halos, in which case clustering alone cannot distinguish them from astrophysical seeds.
\end{abstract}

\maketitle

\section{Introduction}\label{sec:intro}

The James Webb Space Telescope (JWST) has opened a new observational window on the high-redshift Universe, revealing unexpectedly abundant and luminous galaxies at early cosmic times~\cite{Labbe:2022ahb,Boylan-Kolchin:2022kae,2024ApJ...972..143C}. Although at least part of this initial tension can be alleviated through revisions to the modelling of galaxy formation and stellar populations~\cite{Lu:2024oli}, JWST has also uncovered populations whose very nature remains unsettled. Among the most enigmatic are the so-called \textit{little red dots} (LRDs): compact sources found predominantly at $z \sim 4$--8, characterised by distinctive V-shaped rest-frame ultraviolet-to-optical spectral energy distributions and, in many cases, broad hydrogen emission lines~\cite{Matthee:2023utn,2024ApJ...964...39G,Kokorev:2024kqm,2025ApJ...986..126K}.

Several interpretations of LRDs have been proposed, including heavily dust-obscured active galactic nuclei (AGN)~\cite{2023ApJ...954L...4K}, compact and extremely dense star-forming systems, and combinations of nuclear and stellar components~\cite{2024ApJ...968....4P,2025ApJ...984..121W}. The extreme compactness of LRDs and their broad Balmer emission lines point towards accreting black holes, whereas weak X-ray~\cite{2024ApJ...974L..26Y} and mid-infrared~\cite{2024ApJ...968....4P,2025ApJ...984..121W} emission and, in some sources, prominent Balmer breaks~\cite{2024ApJ...964...39G,2025ApJ...995..118S} have motivated alternatives. Among the more exotic interpretations are supermassive stars, whose spectra can reproduce several of the observed ultraviolet-to-optical features without an AGN~\cite{2026ApJ...998..124N,Zwick:2025eik,Lu:2026pif,Nandal:2026chi} and which may form through runaway collisions in the extremely dense stellar cores implied by LRD compactness~\cite{Pacucci:2025ojp}, and quasi-stars, in which a black hole accretes inside a radiation-pressure-supported gaseous envelope~\cite{Begelman:2007je,Santarelli:2025rpm,Hassan:2025qs,Begelman:2025upi,Gentile:2026thl}. Recent spectroscopic observations and modelling, however, increasingly favour scenarios in which a substantial fraction of LRDs are powered by accreting black holes embedded in dense, optically thick gaseous envelopes---so-called \textit{black-hole stars}---which can reproduce several characteristic spectral features~\cite{2026MNRAS.551g1567D,Sun:2026snb}. In this picture, the dense gas can account for the X-ray weakness and Balmer breaks of individual sources~\cite{2024arXiv241204557L}, while electron scattering can explain the broad line profiles~\cite{2026Natur.649..574R}. The black-hole-star model is now being tested against detailed spectroscopy of individual sources~\cite{Merida:2026bhs,Ji:2026bhs}. Related compact, ultraviolet-bright sources, the \textit{little blue dots}, may be connected to LRDs, possibly representing similar systems viewed from different directions or in different accretion states~\cite{Madau:2026ekq,Madau:2026ldo}. The inferred black-hole masses nevertheless remain uncertain: some analyses find black holes that are strongly overmassive relative to their hosts~\cite{Durodola:2024bom,2026Natur.653.1017J}, whereas others attribute the apparent overmassiveness to super-Eddington accretion, implying substantially lower black-hole masses~\cite{Curtis:2026tea,Sun:2026bwh}, of order $10^{4}$--$10^{5}\,M_\odot$ for typical sources, consistent with seeds formed from single supermassive stars~\cite{Sun:2026bwh}.

If accreting black holes power a substantial fraction of LRDs, a central question is how these black holes formed and reached their inferred masses by $z \sim 4$--8. Conventional scenarios invoke seeds from Population~III stellar remnants or direct collapse, followed by efficient accretion~\cite{Madau:2001sc,Maiolino:2023bpi,Jeon:2025dcbh,Cenci:2025lrd,Pacucci:2026ovn,Whalen:2026lrd}, with LRDs possibly marking the first accretion episodes of newly formed seeds~\cite{2025ApJ...988L..22I}. A more exotic alternative is that some of these black holes originated as primordial black holes (PBHs). Enhanced small-scale primordial fluctuations could also shift the formation of the first stars to $z\gtrsim100$; at $z\gtrsim500$, where molecular-hydrogen formation is suppressed by the cosmic microwave background, supermassive stars of $\sim10^{5}\,M_\odot$ could form through atomic cooling alone, providing another channel for very early-forming massive seeds~\cite{Ito:2024mla}. Distinguishing between these possibilities is challenging, because the observed emission primarily probes the present accretion state rather than the formation mechanism of the underlying black hole.

In this work we approach this question using two population-level observables: the comoving number density and the spatial clustering of LRDs~\cite{Pizzati:2024prz}. Their abundance constrains the number of potential host halos or seeds, while their clustering carries information about the environments in which they reside; together they test whether a given formation scenario can consistently account for the observed population. We compare a conventional scenario, in which LRDs occupy virialized dark-matter halos, with one in which they are associated with a dynamically subdominant population of PBHs, and identify the conditions under which a PBH-seed interpretation could remain viable. Since the PBH argument relies only on the seeds being early-forming and dynamically subdominant, our conclusions apply more broadly to any such seed population.

The paper is organised as follows. Section~\ref{sec:observables} introduces the observational constraints on LRD abundance and clustering. Sections~\ref{sec:halos} and~\ref{sec:pbh} examine the halo scenario and the PBH-seed scenario, respectively. Section~\ref{sec:conclusions} discusses the results and presents our conclusions.

Throughout this work we adopt a spatially flat $\Lambda$CDM cosmology with baryon and total matter density parameters $\Omega_b = 0.05$ and $\Omega_m = 0.3$, reduced Hubble parameter $h = 0.7$, and matter power-spectrum normalisation $\sigma_8 = 0.8$. All distances are comoving unless stated otherwise.

\section{Observables}\label{sec:observables}

The two population-level observables that we use to constrain the nature of LRDs are their comoving number density and their spatial clustering strength.

Photometric selections in JWST blank-field surveys---based on the characteristic V-shaped rest-frame UV-to-optical spectral energy distribution---have yielded catalogues of several hundred LRD candidates over $4 \lesssim z \lesssim 9$~\cite{Kokorev:2024kqm,2025ApJ...986..126K}. Spectroscopic follow-up confirms broad Balmer emission lines in a substantial fraction of these sources, supporting the AGN interpretation and providing robust redshifts~\cite{Matthee:2023utn,2024ApJ...964...39G,2023ApJ...954L...4K}. The inferred comoving number density carries factor-of-a-few uncertainties owing to differences in selection criteria, survey depth and area, and the assumed fraction of active sources. Nevertheless, independent analyses consistently find
\be
\bar{n}_{\rm LRD} \sim 10^{-5}\text{--}10^{-4}~\perMpccube
\ee
over the redshift interval $z \approx 4$--8, with no compelling evidence for strong evolution within this range~\cite{Kokorev:2024kqm,Matthee:2023utn,2025ApJ...986..126K,2024ApJ...964...39G,Zhuang:2025nexus,Pan:2026nexus}. For the purposes of this work we therefore treat the LRD number density as approximately constant over $z = 4$--8 and adopt the above range as our observational constraint.

Measurements of LRD spatial clustering are still in their early stages, limited by the modest sample sizes available to date. Cross-correlation and angular clustering analyses indicate, however, that LRDs are strongly biased tracers of large-scale structure, with effective linear bias values in the range $b \sim 3$--8 over $z \approx 4$--8, higher values being suggested at higher redshifts~\cite{Zebrowski:2026nsk,Wang:2026vzg}. Cross-correlation analyses of 98 spectroscopically confirmed LRDs at $3<z<7$ indicate host-halo masses growing from $\lesssim10^{10.1}\,M_\odot$ at $z\sim7.5$ to $\sim10^{11.3}\,M_\odot$ at $z\sim3.5$~\cite{Zhang:2026beu}, corresponding to $b\lesssim5$ and $b\approx3$, respectively.
Such an increase with redshift, if confirmed, would be consistent with the expectation that massive black-hole seeds occupied progressively rarer and more strongly biased density peaks at earlier cosmic times. Excess clustering on kiloparsec scales, in the form of a high incidence of close LRD pairs, has also been reported~\cite{Tanaka:2024lrd}. Not all analyses find consistent evidence for enhanced large-scale clustering; some studies do not detect a significant excess of LRDs in overdense environments at fixed redshift~\cite{Carranza:2025lrd}, which may reflect differences in selection criteria or the angular scales probed. We therefore adopt $b \simeq 3$--8 as a representative range encompassing current clustering estimates over $z = 4$--8, rather than as a precisely measured redshift evolution, and use it as the primary observational constraint on clustering strength.

\section{Purely adiabatic fluctuations. Formation at virialized density peaks}\label{sec:halos}

\begin{figure}
  \begin{center}
    \includegraphics[width=\columnwidth]{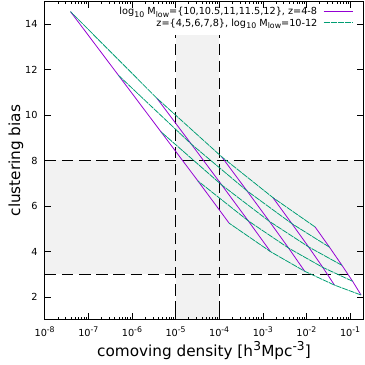}
  \end{center}
  \caption{Effective clustering bias $\bar{b}$ versus comoving number density $\bar{n}$ of dark-matter halos computed with the SMT model~\cite{Sheth:1999su}. Solid purple lines show constant lower mass cutoffs $\log_{10}(M_{\rm low}/\hMsun) \in \{10,10.5,11,11.5,12\}$ varying redshift over $z=4$--8; dashed teal lines show constant redshifts $z\in\{4,5,6,7,8\}$ varying $M_{\rm low}$. Grey shaded bands mark the observational constraints on $\bar{n}_{\rm LRD}$ (vertical) and $\bar{b}$ (horizontal).}
  \label{fig:halo_grid}
\end{figure}

Within the standard $\Lambda$CDM framework, the natural formation sites for the massive black holes powering LRDs are dark-matter halos. We adopt the simplest possible model: every dark-matter halo above a minimum mass $M_{\rm low}$ hosts a single supermassive black hole, which is observed as an LRD with probability $f_{\rm on} \leq 1$. The parameter $f_{\rm on}$ can be interpreted as an effective duty cycle, encoding the fraction of time the AGN is active, any orientation- or obscuration-dependent selection effects, or other sources of incompleteness between the underlying halo population and the observed LRD sample. The model thus contains only one parameter relevant for the clustering prediction, $M_{\rm low}$, while $f_{\rm on}$ merely fixes the abundance. Despite its simplicity, it is sufficient for the comparative analysis we carry out here.

Under these assumptions, the mean comoving LRD number density and effective clustering bias are

\be
\bar n(z)=f_{\rm on}\int_{M_{\rm low}}\dd M\,\frac{\dd n}{\dd M}(M,z)
\ee

\be
\bar b(z)=\frac{\int_{M_{\rm low}}\dd M\,\frac{\dd n}{\dd M}(M,z)\,b(M,z)}{\int_{M_{\rm low}}\dd M\,\frac{\dd n}{\dd M}(M,z)}\,,
\ee

where $\dd n/\dd M$ is the halo mass function and $b(M,z)$ is the linear halo bias. Note that $\bar b$ is independent of $f_{\rm on}$, so the duty cycle affects only the number density. For both the mass function and the bias we adopt the Sheth--Mo--Tormen (SMT) fitting formulae~\cite{Sheth:1999su}.

The results of this calculation are shown in Fig.~\ref{fig:halo_grid}. The two families of curves trace, respectively, the locus of constant lower mass cutoff (varying redshift) and constant redshift (varying $M_{\rm low}$) in the $(\bar{n}, \bar{b})$ plane. The grey shaded bands mark the observational constraints on number density and bias described in Sect.~\ref{sec:observables}. 

\begin{figure}
  \begin{center}
    \includegraphics[width=\columnwidth]{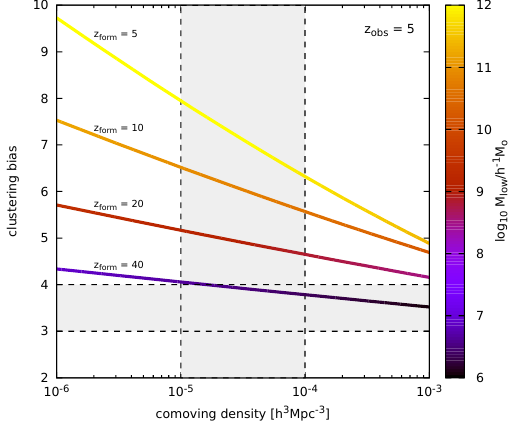}
  \end{center}
  \caption{Bias evolution of a tracer population formed at redshift $z_f > z_{\rm obs}$, following Eq.~(\ref{eq:fry})~\cite{Fry:1996fg}. For sources formed at $z_f \gtrsim 10$, the effective bias at the observed epoch $z_{\rm obs} \sim 5$ is substantially reduced relative to the instantaneous-formation prediction, enabling the lower end of the observed LRD bias range to be accommodated within the halo model.}
  \label{fig:highz_form}
\end{figure}

Since $\bar{b}$ is independent of $f_{\rm on}$, it is the bias measurement that directly constrains $M_{\rm low}$, while the number density then determines $f_{\rm on}$. The observed bias range $b \sim 3$--8 over $z = 4$--8 is reproduced for $M_{\rm low} \sim 10^{10.5}$--$10^{12}~\hMsun$. At the lower end of this range the cumulative halo number density substantially exceeds the observed LRD abundance, requiring $f_{\rm on} \ll 1$ to bring the two into agreement. For $M_{\rm low} \gtrsim 10^{11.5}~\hMsun$ the cumulative number density falls within the observed range, permitting $f_{\rm on}$ of order unity. In all cases, reducing $f_{\rm on}$ leaves $\bar{b}$ unchanged, so the bias and number density constraints can always be satisfied simultaneously. One way to reduce $\bar b$ while keeping $\bar n$ fixed is to consider a scenario in which LRDs formed at an earlier epoch $z_f > z_{\rm obs}$ and have since evolved passively in the potential landscape of the dominant CDM field. Following Fry~\cite{Fry:1996fg}, the bias of a tracer population formed at $z_f$ with bias $b_f$ evolves as
\be\label{eq:fry}
b(z_{\rm obs}) = 1 + \frac{D(z_f)}{D(z_{\rm obs})}\,(b_f - 1)\,,
\ee
where $D(z)$ is the linear growth factor. Since $D(z_f)/D(z_{\rm obs}) < 1$ for $z_f > z_{\rm obs}$, the bias decays toward unity as the Universe expands. This mechanism is illustrated in Fig.~\ref{fig:highz_form}: LRDs that formed at sufficiently high redshift ($z_f \gtrsim 10$) can exhibit a substantially reduced bias at $z_{\rm obs} \sim 5$, providing a possible route to the lower end of the observed bias range. Equation~(\ref{eq:fry}) assumes passive evolution of a conserved tracer population, with no further formation, mergers or destruction. LRDs, which most likely represent transient AGN-like phases of black holes that may continue to grow and merge together with their host halos, are not obviously conserved tracers, so Eq.~(\ref{eq:fry}) should be regarded as an illustrative limiting case rather than a realistic model of LRD bias evolution. In contrast, it is well suited to dynamically subdominant PBHs (Sect.~\ref{sec:pbh}): at the tiny abundances required here, with mean comoving separations of $\bar n_{\rm pbh}^{-1/3}\sim 20$--$50~h^{-1}\,{\rm Mpc}$, PBH mergers are negligible, so PBHs are effectively conserved as a population and simply follow the dominant CDM potential.

Taken together, these results show that the simple halo model naturally accommodates both the observed number density and clustering strength of LRDs over $z = 4$--8, without fine-tuning. A more sophisticated halo occupation distribution (HOD) approach~\cite{Cooray:2002dia} would allow additional flexibility (scatter in the occupation number, a central-satellite decomposition, etc.), but the current observational precision does not motivate the additional parameters. This analysis concerns the probable \textit{formation sites} of LRDs; the detailed physical \textit{formation mechanisms} of the black holes and the conditions giving rise to AGN activity remain active areas of research (see Sect.~\ref{sec:intro} and references therein).

\section{Additional isocurvature component. Primordial black holes}\label{sec:pbh}

\begin{figure}
  \begin{center}
    \includegraphics[width=\columnwidth]{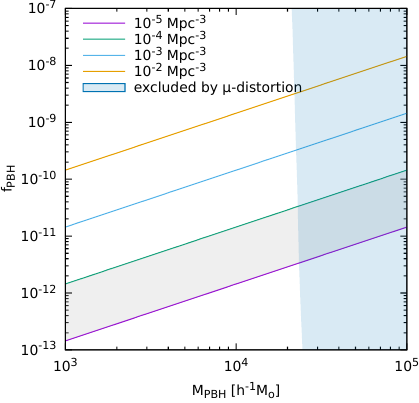}
  \end{center}
  \caption{Required PBH mass fraction $f_{\rm pbh}$ as a function of PBH mass $M_{\rm pbh}$ to match the observed LRD number density (grey shaded band), assuming $f_{\rm on}=1$; for $f_{\rm on}<1$ the band shifts upward by a factor $f_{\rm on}^{-1}$ [Eq.~(\ref{eq:fpbh})]. The blue shaded region is excluded by CMB spectral $\mu$-distortions for Gaussian initial fluctuations. For all masses considered, the required $f_{\rm pbh}$ is negligibly small, confirming that PBHs in this regime are dynamically unimportant for large-scale structure.}
  \label{fig:pbh_ndens}
\end{figure}

Since the formation of supermassive black hole seeds at $z \sim 4$--8 remains challenging in the standard $\Lambda$CDM framework---requiring either sustained near-Eddington accretion from low-mass stellar remnants or the existence of anomalously massive early seeds---alternative seeding channels have attracted considerable attention. Among the most widely discussed alternatives are primordial black holes (PBHs): compact objects formed in the early Universe through the gravitational collapse of over-dense regions in the primordial plasma~\cite{Zeldovich:1967lct,Hawking:1971ei,Carr:1974nx,Carr:1975qj,Meszaros:1975ef}.
PBHs have long been discussed as a source of enhanced early structure formation~\cite{Meszaros:1975ef,1977A&A....56..377C,1983ApJ...268....1C}. Their discreteness enhances small-scale power both through Poisson fluctuations in the PBH number density and through the seed effect around individual PBHs, as quantified in more recent analytic and numerical studies~\cite{Afshordi:2003zb,Carr:2018rid,Hutsi:2019hlw,Inman:2019wvr,Cappelluti:2021usg,Delos:2024poq,Zhang:2024ytf,Magaraggia:2026ngp}.
More specifically, PBHs have been proposed as seeds for the supermassive black holes powering high-redshift AGN and as a possible explanation for the excess of bright, massive galaxies at $z \gtrsim 10$ revealed by JWST~\cite{Liu:2022bvr,Hutsi:2022fzw,Huang:2023mwy,Yuan:2023bvh,Su:2023jno,Gouttenoire:2023nzr,Dolgov:2023ijt,Dayal:2024zwq,Liu:2023pvq,Colazo:2024jmz,Huang:2024koy,Huang:2024aog,Ziparo:2024nwh,Matteri:2025klg,Dayal:2025aiv,Prole:2025snf,Colazo:2025xec,Koulen:2025xjq,Ivanov:2025pbu,Li:2025jxq,Kashlinsky:2026lci,Chatterjee:2026nmb}.
Several recent works have proposed that PBHs could act as seeds for the black holes observed in LRDs~\cite{Hai-LongHuang:2024gtx,DeLuca:2025nao,Guo:2026cuv,Zhang:2025asq,Zhang:2025grn,Zhang:2025oyl,Zhang:2026fhn,Maiolino:2025tih}, and whether their clustering properties are consistent with observations has begun to be addressed~\cite{Zhang:2025tgm,Zhang:2026uum,Wang:2026vzg}. In particular, Ref.~\cite{DeLuca:2025nao} explicitly investigated PBH-based pathways to LRDs: direct formation of PBHs with the inferred LRD black-hole masses is excluded by CMB $\mu$-distortion limits, assembly from lighter PBHs via hierarchical mergers is hampered by the rarity of the required high-redshift host halos, whereas gas accretion onto intermediate-mass PBHs can reproduce the observed LRD properties in parts of parameter space.
PBHs represent one of the oldest dark matter candidates~\cite{Chapline:1975ojl}, and they remain observationally allowed over a broad range of masses. For most of this range, however, gravitational microlensing, gravitational-wave observations, and accretion signatures already exclude PBHs as the dominant dark matter component, with the only remaining open window at asteroid masses ($10^{-17}$--$10^{-10}\,M_\odot$)~\cite{Carr:2020gox,Carr:2026hot}. At the masses relevant for SMBH seeding ($10^{3}$--$10^{5}\,\hMsun$), PBHs can constitute only a tiny fraction of the cold dark matter.

The mean comoving number density of PBHs of mass $M_{\rm pbh}$ that constitute a fraction $f_{\rm pbh} \equiv \Omega_{\rm pbh}/\Omega_{\rm cdm}$ of the CDM is $\bar n_{\rm pbh} = f_{\rm pbh}\,\Omega_{\rm cdm}\,\rho_{\rm crit}/M_{\rm pbh}$, where $\rho_{\rm crit}$ is the present-day critical density. If only a fraction $f_{\rm on}\leq1$ of PBHs is observed as LRDs at any given time (cf.\ Sect.~\ref{sec:halos}), then $\bar n_{\rm LRD}=f_{\rm on}\,\bar n_{\rm pbh}$, and the required PBH mass fraction is
\begin{multline}\label{eq:fpbh}
f_{\rm pbh}\simeq1.4\times10^{-10}f_{\rm on}^{-1}\left(\frac{\bar n_{\rm LRD}}{10^{-4}\,\perMpccube}\right)\\
\times\left(\frac{M_{\rm pbh}}{10^5~\hMsun}\right)\left(\frac{\Omega_{\rm cdm}}{0.25}\right)^{-1}\,.
\end{multline}
The combinations of $f_{\rm pbh}$ and $M_{\rm pbh}$ required to reproduce the observed LRD number density for $f_{\rm on}=1$ are shown in Fig.~\ref{fig:pbh_ndens}. For PBH masses in the range $10^{3}$--$10^{5}~\hMsun$, matching $\bar{n}_{\rm LRD} \sim 10^{-5}$--$10^{-4}~\perMpccube$ requires $f_{\rm pbh} \sim 10^{-11}$--$10^{-10}$ for $M_{\rm pbh} = 10^5~\hMsun$, falling to $f_{\rm pbh} \sim 10^{-13}$--$10^{-12}$ for $M_{\rm pbh} = 10^3~\hMsun$ (since $\bar{n}_{\rm pbh} \propto f_{\rm pbh}/M_{\rm pbh}$, lighter PBHs require a smaller mass fraction to reach the same number density). Across the full mass range the required values span $f_{\rm pbh} \sim 10^{-13}$--$10^{-10}$: an extremely small CDM fraction. Also shown in Fig.~\ref{fig:pbh_ndens} are upper bounds on $f_{\rm pbh}$ from CMB spectral $\mu$-distortions~\cite{Zeldovich:1969ff,1970CoASP...2...66S,Chluba:2011hw,Chluba:2012we}, which arise from the Silk damping~\cite{Silk:1967kq} of small-scale perturbations required to produce PBHs under the assumption of Gaussian initial fluctuations~\cite{Nakama:2017xvq}. These bounds can in principle be circumvented in non-Gaussian scenarios~\cite{Byrnes:2024vjt,Allegrini:2026jqt,Dave:2026gsd} or for alternative PBH formation mechanisms, although they remain largely robust even for large cubic ($g_{\rm NL}$) non-Gaussianity~\cite{Pritchard:2025yda}. Non-Gaussian PBH seeds of supermassive black holes are further constrained by dark-matter substructure~\cite{Ando:2024ghr} and can be tested with future spectral-distortion and pulsar-timing observations~\cite{Unal:2020mts}. In all cases, however, the required $f_{\rm pbh}$ values place PBHs far below the dynamical threshold for any significant effect on large-scale structure. For $f_{\rm on} < 1$ (\ie{} not all PBHs shine as LRDs), $f_{\rm pbh}$ must be increased by a factor $f_{\rm on}^{-1}$ [Eq.~(\ref{eq:fpbh})], but unless $f_{\rm on}$ is extremely small the qualitative conclusion is unchanged: PBHs in this mass range are dynamically negligible.

A seed PBH of initial mass $\sim 10^4~\hMsun$ can grow by roughly an order of magnitude through accretion of primordial gas by $z \sim 5$~\cite{Ali-Haimoud:2016mbv}; the growth is more restricted for $10^3~\hMsun$ seeds and can be considerably larger for $10^5~\hMsun$ seeds. This mass range is also independently motivated by recent spectral analyses of LRDs, which attribute their apparent overmassiveness to super-Eddington accretion~\cite{Curtis:2026tea,Sun:2026bwh} and infer typical black-hole masses of $\sim 10^{4}$--$10^{5}~M_\odot$, consistent with seeds as massive as single supermassive stars~\cite{Sun:2026bwh}. In the following we restrict our analysis to the range $M_{\rm pbh} = 10^{3}$--$10^{5}~\hMsun$.

\begin{figure}
  \begin{center}
    \includegraphics[width=\columnwidth]{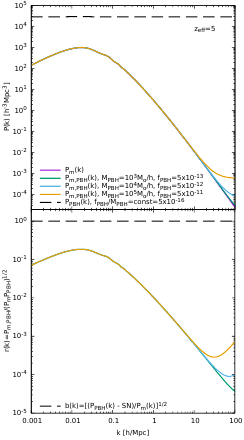}
  \end{center}
  \caption{Power spectra at $z = 5$ for three combinations of $M_{\rm pbh}$ and $f_{\rm pbh}$ chosen to reproduce $\bar{n}_{\rm pbh} \simeq 4\times10^{-5}~\perMpccube$. \textit{Upper panel:} total matter spectrum $P_{\rm m}$ (purple), matter-PBH cross-spectra $P_{\rm m,pbh}$ (green, blue, orange), and PBH auto-spectrum $P_{\rm pbh}$ (dashed). The PBH auto-spectrum is dominated by Poisson noise, and the effect of PBHs on $P_{\rm m}$ is negligible. \textit{Lower panel:} matter-PBH cross-correlation coefficient $r$ (solid) and Poisson-subtracted PBH bias $b_{\rm pbh}$ (dashed). After shot-noise subtraction $b_{\rm pbh} \simeq 1$, \ie{} PBHs are unbiased tracers of the adiabatic density field. The low values of $r$ reflect the dominance of Poisson noise, $r \simeq [\bar n_{\rm pbh}P_{\rm m}/(1+\bar n_{\rm pbh}P_{\rm m})]^{1/2}$; the isocurvature contribution is visible only as the small-scale upturn for the most massive PBHs.}
  \label{fig:pk_pbh}
\end{figure}

To describe the evolution of large-scale structure in the presence of a subdominant PBH component, we follow the formalism of Inman \& Ali-Ha\"imoud~\cite{Inman:2019wvr},  extended minimally to capture the matter-PBH cross-power spectrum. The three relevant power spectra --- the total matter spectrum $P_{\rm m}$, the PBH auto-spectrum $P_{\rm pbh}$, and the matter-PBH cross-spectrum $P_{\rm m,pbh}$ --- are

\bea
P_{\rm m}(k,z)&=&P_{\rm m}^{\rm ad}(k,z)+D_{\rm iso}^2(z)f_{\rm pbh}^2\,\bar n_{\rm pbh}^{-1}\label{eq:Pm}\\
P_{\rm pbh}(k,z)&=&P_{\rm m}^{\rm ad}(k,z)+\nonumber\\
&+&\left[1+ (D_{\rm iso}(z)-1)f_{\rm pbh}\right]^2\,\bar n_{\rm pbh}^{-1}\label{eq:Ppbh}\\
P_{\rm m,pbh}(k,z)&=&P_{\rm m}^{\rm ad}(k,z)+D_{\rm iso}(z)f_{\rm pbh}\cdot\nonumber\\
&\cdot&\left[1+ (D_{\rm iso}(z)-1)f_{\rm pbh}\right]\,\bar n_{\rm pbh}^{-1}\label{eq:Pmpbh}
\eea

Here $P_{\rm m}^{\rm ad}$ is the standard adiabatic matter power spectrum and $\bar{n}_{\rm pbh}^{-1}$ is the Poisson shot-noise contribution from the discrete PBH distribution. The isocurvature growth factor $D_{\rm iso}(z)$ describes how the local PBH overdensity --- initially uncorrelated with the adiabatic matter field (a pure isocurvature mode) --- develops a coherent component as gravitational instability proceeds. In terms of these spectra we define the PBH clustering bias and the matter-PBH cross-correlation coefficient as

\be\label{eq:bpbh}
b^2_{\rm pbh}(k,z)\equiv\frac{P_{\rm pbh}(k,z)-{\rm SN}}{P_{\rm m}(k,z)}
\ee

\be\label{eq:r}
r(k,z)\equiv\frac{P_{\rm m,pbh}(k,z)}{\sqrt{P_{\rm m}(k,z)P_{\rm pbh}(k,z)}}\,,
\ee
where ${\rm SN}\equiv\bar n_{\rm pbh}^{-1}$ is the constant Poisson shot-noise contribution; its subtraction makes $b_{\rm pbh}$ directly comparable to the observed tracer biases, which are shot-noise subtracted in the same way.\footnote{If cross-spectra are available, the PBH bias can alternatively be defined as $b_{\rm pbh}^{\times}(k,z)\equiv P_{\rm m,pbh}(k,z)/P_{\rm m}(k,z)$, which is free of the auto-spectrum shot noise. For the model of Eqs.~(\ref{eq:Pm})--(\ref{eq:Pmpbh}) both definitions give $b_{\rm pbh}\simeq1$ on linear scales. Since the matter field is not directly observable, the corresponding observational estimator uses cross-correlation with a tracer of known bias, as in the LRD cross-correlation analyses of Refs.~\cite{Zebrowski:2026nsk,Wang:2026vzg}.} The isocurvature term in Eq.~(\ref{eq:Ppbh}) in excess of SN represents genuine clustering induced by the discrete PBH distribution and is retained.

The approximation used for $D_{\rm iso}$ is taken from Ref.~\cite{Inman:2019wvr}:

\bea
D_{\rm iso}(z)&\simeq&\left(1+\frac{3\gamma}{2\alpha_-}s\right)^{\alpha_-}\label{eq:Diso}\\
s&\equiv&\frac{1+z_{\rm eq}}{1+z}\nonumber\\
\alpha_-&\equiv&\frac{1}{4}\left(\sqrt{1+24\gamma}-1\right)\nonumber\\
\gamma&\equiv&1-f_{\rm b}\nonumber
\eea

Here $z_{\rm eq}$ is redshift for matter-radiation equality and $f_{\rm b}\equiv \Omega_{\rm b}/\Omega_{\rm m}$ is the baryonic fraction. This approximation assumes matter domination and thus neglects the late-time suppression of growth by $\Lambda$, which is unimportant at the LRD redshifts $z \gtrsim 4$.

Example power spectra at $z_{\rm obs} = 5$ are shown in Fig.~\ref{fig:pk_pbh} for values of $M_{\rm pbh}$ and $f_{\rm pbh}$ that reproduce a number density $\bar{n}_{\rm pbh} \simeq 4 \times 10^{-5}~\perMpccube$, comparable to the observed LRD abundance. The PBH auto-spectrum is dominated by Poisson shot noise and is well approximated by a white-noise floor. The impact of PBHs on the total matter power spectrum is negligibly small, consistent with their tiny mass fraction. The matter-PBH cross-spectrum is less affected by discreteness noise and therefore provides the cleanest handle on the clustering of PBHs relative to matter. The lower panel of Fig.~\ref{fig:pk_pbh} shows the cross-correlation coefficient $r(k)$ and the Poisson-subtracted PBH bias. After subtracting the shot-noise contribution, the residual PBH bias is $b_{\rm pbh} \simeq 1$: since the PBH component is dynamically negligible and initially Poisson distributed, PBHs simply trace the adiabatic matter field. The cross-correlation coefficient is small, $r \lesssim 0.2$ on linear scales at $z = 5$, but this is almost entirely due to Poisson noise, $r \simeq [\bar n_{\rm pbh}P_{\rm m}/(1+\bar n_{\rm pbh}P_{\rm m})]^{1/2}$, rather than to the isocurvature nature of the PBH perturbations, whose contribution [Eqs.~(\ref{eq:Pm})--(\ref{eq:Pmpbh})] becomes appreciable only at $k \gtrsim 10\,h\,\mathrm{Mpc}^{-1}$ for the most massive PBHs, where nonlinear evolution is expected to become important and our linear treatment is no longer quantitatively reliable. The decisive quantity is therefore the bias: $b_{\rm pbh} \simeq 1$, compared with the observed $b \sim 3$--8. Note that a constant duty cycle $f_{\rm on}<1$ does not help: if LRDs are a random subsample of the PBH population, their bias is unchanged, $b_{\rm LRD}=b_{\rm pbh}\simeq1$, and only the shot noise increases to $\bar n_{\rm LRD}^{-1}=(f_{\rm on}\bar n_{\rm pbh})^{-1}$.

The essential difficulty for the PBH scenario --- and, more broadly, for any picture in which LRDs are very early-forming compact seeds --- is the predicted clustering bias. Because PBHs are dynamically subdominant, they follow the potential landscape generated by the dominant adiabatic CDM fluctuations, and their bias evolves according to Eq.~(\ref{eq:fry}), asymptotically approaching $b \rightarrow 1$ on a Hubble timescale. Unless the initial PBH clustering is very finely tuned, the effective bias at $z \sim 5$ will be of order unity --- far below the observed values of $b \sim 3$--8. The same argument applies to any population of very early-forming seeds that is dynamically unimportant and whose activity is not tied to massive host halos (see Sect.~\ref{sec:conclusions}). This should be contrasted with the standard halo scenario, where the observed bias range is reached naturally and without fine-tuning.

\section{Discussion and Conclusions}\label{sec:conclusions}

We have used the observed comoving number density and spatial clustering of JWST little red dots over $z = 4$--8 to compare two broad scenarios for their origin. In the standard adiabatic scenario, LRDs form at virialized dark-matter density peaks. In the alternative isocurvature scenario, they are associated with a dynamically subdominant population of primordial black holes. Our main findings are as follows.

The halo model provides a natural and self-consistent account of both observables. A simple model in which every dark-matter halo above a mass threshold $M_{\rm low}$ hosts an LRD with duty cycle $f_{\rm on}$ reproduces the observed bias range $b \sim 3$--8 across $z = 4$--8 for $M_{\rm low} \sim 10^{10.5}$--$10^{12}~\hMsun$, while the observed number density $\bar{n}_{\rm LRD} \sim 10^{-5}$--$10^{-4}~\perMpccube$ is then matched by an appropriate duty cycle, without fine-tuning. Bias values at the lower end of the observed range can also be accommodated, in the limiting case of passive evolution, by sources formed at earlier epochs. Extending the model to a full HOD framework~\cite{Cooray:2002dia} is possible but not motivated by current data.

The PBH scenario faces a fundamental challenge. Matching the observed LRD number density requires $f_{\rm pbh} \sim 10^{-13}$--$10^{-10}$ across $M_{\rm pbh} = 10^3$--$10^5~\hMsun$, making PBHs entirely negligible for structure formation. For such a dynamically subdominant Poisson distribution of seeds, the power spectrum analysis of Sect.~\ref{sec:pbh} shows directly that $b_{\rm pbh} \simeq 1$ at $z \sim 5$ --- far below the observed LRD clustering signal. More broadly, the same conclusion applies to any population of very early-forming seeds with initially unclustered (Poisson) statistics that is dynamically unimportant. If instead such seeds were initially clustered, the Fry equation~(\ref{eq:fry}) governs the subsequent bias evolution: being dynamically subdominant, they follow the adiabatic CDM potential and their bias decays toward unity, so that without fine-tuning of the initial clustering, $b \sim 1$ is again the generic prediction at $z \sim 5$.

Non-Gaussian scenarios for PBH formation could in principle provide such initial clustering~\cite{Khlopov:2004sc,Clesse:2015wea}. Strongly non-Gaussian initial perturbations could both bypass the $\mu$-distortion bounds~\cite{Byrnes:2024vjt,Allegrini:2026jqt,Dave:2026gsd} (but see Refs.~\cite{Pritchard:2025yda,Ando:2024ghr}) and produce enhanced initial PBH clustering~\cite{Zhang:2025tgm,Zhang:2026uum}. However, generating $b \sim 5$--8 at $z \sim 5$ from PBH clustering alone would require a level of primordial non-Gaussianity that is difficult to motivate without fine-tuning.

Another possible way to reconcile PBH seeds with the observed clustering is an environment-dependent activation probability, $f_{\rm on}=f_{\rm on}(M_{\rm h})$, such that only PBHs residing in sufficiently massive host halos accrete efficiently enough to appear as LRDs. Such a coupling is physically plausible: the accretion rate is set by the local gas density and temperature, which are strongly enhanced in the centres of massive halos, where dynamical friction also brings the most massive black holes. Conversely, in less massive halos supernova feedback efficiently expels gas from the central regions and stunts black-hole growth, so that efficient accretion sets in only above a critical host-halo mass~\cite{2015MNRAS.452.1502D,2017MNRAS.468.3935H,2017MNRAS.465...32B,Angles-Alcazar:2017rvx}. The accretion-driven growth of intermediate-mass PBHs considered in Ref.~\cite{DeLuca:2025nao} relies on an efficient gas supply, which naturally points to such massive host environments. Since PBHs trace the matter distribution, the fraction residing in halos above $M_{\rm low}\sim10^{10.5}$--$10^{11.5}~\hMsun$ at $z\sim5$, and hence the effective $f_{\rm on}$, is only $\sim10^{-3}$--$10^{-2}$, so that by Eq.~(\ref{eq:fpbh}) the required PBH abundance increases correspondingly to $f_{\rm pbh}\sim10^{-11}$--$10^{-7}$, still dynamically negligible. In this case, however, the clustering of LRDs is determined entirely by their host halos, and the scenario becomes observationally indistinguishable from the halo picture of Sect.~\ref{sec:halos}: the PBH merely provides the seed, while both the location and the activity of the LRD are governed by the halo. The clustering measurements therefore rule out PBHs (or similar early-forming seeds) as a randomly activated, unbiased LRD population, but cannot by themselves exclude a primordial origin of black holes that subsequently grow in massive halos. Discriminating between primordial and astrophysical seeds in this case requires other observables, such as the seed mass function at $z\gtrsim10$, the black-hole occupation fraction of low-mass halos, or gravitational-wave signatures. Gravitational-wave observations are particularly promising in this respect. The LISA mission will detect mergers of massive black holes with masses $\sim10^4$--$10^7\,M_\odot$ out to $z\sim20$, and the inferred mass and redshift distributions of these events are sensitive to the seeding channel~\cite{Sesana:2007sh,Klein:2015hvg,LISA:2017pwj,Ricarte:2018mzn,Ellis:2023owy}. Proposed mid-band atom interferometers, such as AEDGE and AION, would extend this sensitivity to intermediate-mass black-hole binaries in the $\sim10^3$--$10^5\,M_\odot$ range directly relevant for LRD seeds~\cite{AEDGE:2019nxb,Badurina:2019hst,Ellis:2023iyb}, while the nanohertz gravitational-wave background reported by pulsar timing arrays provides complementary constraints on the growth history of the most massive black holes, including those observed by JWST~\cite{Ellis:2023dgf,Ellis:2024wdh,Urrutia:2024hwc}. Third-generation ground-based detectors, such as the Einstein Telescope, could in addition detect black-hole mergers at $z\gtrsim30$, before the onset of significant star formation, which would provide a strong indication of a primordial origin for at least part of the black-hole population~\cite{Koushiappas:2017kqm,ET:2019dnz,Ng:2021sqn,ET:2025xjr}; more generally, future detectors can probe PBH binaries over a broad mass range, building on detailed modelling of their formation and merger rates~\cite{Raidal:2017mfl,Raidal:2018bbj,Vaskonen:2019jpv,Pujolas:2021yaw,Raidal:2024bmm}.

Observational evidence for a close connection between LRDs and overdense large-scale environments provides additional qualitative support for the halo picture. LRDs appear preferentially associated with bright companion galaxies and enhanced galaxy overdensities~\cite{Pan:2026nexus}, consistent with the expectation that black-hole growth is facilitated by the dense gas reservoirs and elevated Lyman-Werner radiation backgrounds characterising biased nodes of the cosmic web. Strikingly, the companions of the brightest LRDs are found to be both more numerous and bluer than those of control galaxies at $>3\sigma$ significance~\cite{Pacucci:2026occ}, consistent with UV-bright neighbours supplying the Lyman-Werner flux required for direct collapse. This environmental correlation supports the direct-collapse black-hole scenario, in which nearby UV sources suppress molecular hydrogen cooling in metal-poor atomic-cooling halos, enabling the formation of massive seeds without fragmentation. The environmental evidence is, however, not yet conclusive: an overdensity analysis of spectroscopically confirmed LRDs finds them predominantly in underdense regions at $z>4$, shifting to typical galaxy environments by $z\sim3.5$~\cite{Zhang:2026beu}.

The Fisher analysis presented in the Appendix shows that for a COSMOS-Web-like survey~\cite{Casey:2022amu}---the widest-area blank-field JWST programme, covering $\sim 0.5$~deg$^2$---over $z = 4$--8, clustering can be detected at $\geq 3\sigma$ only for bias values $b \gtrsim 3$--5, depending on the LRD number density (Fig.~\ref{fig:snr}), fully consistent with the clustering strengths reported in the literature. The standard PBH prediction of $b \simeq 1$ is then ruled out at approximately $(b_{\rm obs} - 1)/\sigma_b = {\rm SNR}(1 - 1/b_{\rm obs})$ sigma, \eg, if clustering of $b = 4$ tracers is detected at $5\sigma$ significance, the $b = 1$ case is disfavoured at $\sim 3.8\sigma$. A survey with ten times the sky coverage would improve the signal-to-noise by a factor $\sqrt{10}\approx 3$, enabling a definitive discrimination between the halo and PBH scenarios.

In summary, the spatial clustering of JWST little red dots is most naturally explained within the standard $\Lambda$CDM framework, where LRDs occupy dark-matter halos with masses $M_{\rm low} \sim 10^{10.5}$--$10^{12}~\hMsun$ at $z = 4$--8. Primordial black holes, and more broadly any population of dynamically subdominant early-forming seeds, predict a clustering bias of order unity, substantially below the observed $b \sim 3$--8, if their activity is not tied to their environment, and are therefore disfavoured by current data as a randomly activated LRD population. If instead only seeds residing in massive halos become LRDs, the clustering is inherited from the host halos and cannot discriminate between primordial and astrophysical seed origins. Future measurements of LRD clustering with wider-field JWST surveys, combined with cross-correlation analyses against galaxy and quasar samples, will provide an increasingly precise test of these conclusions, while upcoming gravitational-wave observations offer a complementary probe of the seed origin itself.

\acknowledgments

This work was supported by the Estonian Research Council grants TARISTU24-TK10 and TARISTU24-TK3 and the Center of Excellence program TK202 ``Foundations of the Universe''. Claude (Anthropic) and ChatGPT (OpenAI) have been used solely as writing assistants.

\appendix
\section{Clustering amplitude. Fisher analysis}

\begin{figure}[!b]
  \begin{center}
    \includegraphics[width=\columnwidth]{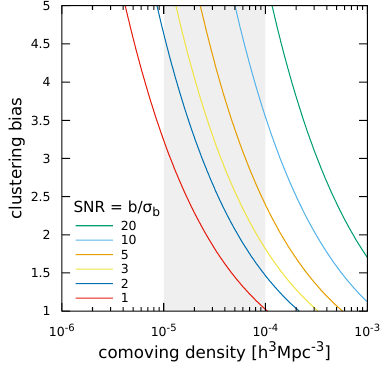}
  \end{center}
  \caption{Signal-to-noise ratio for the clustering amplitude as a function of clustering bias and tracer number density, for a COSMOS-Web-like survey~\cite{Casey:2022amu} covering $\sim 0.5$~deg$^2$ over $z = 4$--8. The grey shaded band shows LRD-like number densities. Clustering can be detected at $\geq 3\sigma$ only for $b \gtrsim 3$--5, depending on $\bar n$, consistent with reported LRD clustering strengths.}
  \label{fig:snr}
\end{figure}

\begin{figure}[!b]
  \begin{center}
    \includegraphics[width=\columnwidth]{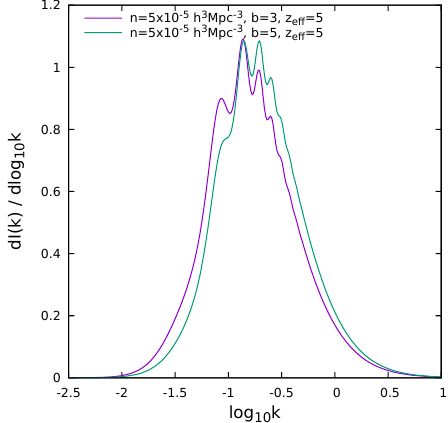}
  \end{center}
  \caption{Contribution per logarithmic $k$-interval to the Fisher integral, $\dd I/\dd\log_{10}k=\ln 10\,k^3[\bar{n}P(k)/(1+\bar{n}P(k))]^2$, at $z_{\rm eff}=5$ for $\bar n=5\times10^{-5}~\perMpccube$ and bias values $b=3$ and 5, illustrating the $k$-range that dominates the signal-to-noise on the clustering amplitude.}
  \label{fig:snr_intgrand}
\end{figure}

Power spectrum Fisher matrix for set of parameters $\{\theta_i\}$ can be given as, \eg~\cite{Tegmark:1997rp}
\benn
F_{ij} = \sum_{k} \frac{1}{\sigma_P^2(k)}\,
\frac{\partial P(k)}{\partial\theta_i}\,
\frac{\partial P(k)}{\partial\theta_j}\,,
\eenn
where the spectrum variance
\benn
\sigma_P^2(k)=\frac{2P(k)^2}{N_k}\,.
\eenn

Here $N_k$ is the effective number of modes
\benn
N_k=\frac{V_{\rm eff}}{(2\pi)^3}V_{\rm bin}(k)\,,
\eenn
$V_{\rm eff}$ is the effective survey volume, which for constant $\bar n$
\benn
V_{\rm eff}=\left[\frac{\bar nP(k)}{1+\bar nP(k)}\right]^2V_{\rm survey}\,,
\eenn
and $V_{\rm bin}(k)=4\pi k^2\Delta k$ is the volume of the k-space shell.

Assuming known cosmology and the only unknown parameter to be the linear clustering bias $b$ ($P(k)=b^2P_{\rm m}(k)$) the variance for $b$ can be expressed as
\benn
\sigma_b^2=\frac{b^2}{2\sum\limits_{k} N_k}=\frac{\pi^2 b^2}{V_{\rm survey}\cdot I(k)}\,,
\eenn
where for continuous case
\benn
I(k)=\int \dd k\, k^2\left[\frac{\bar nP(k)}{1+\bar nP(k)}\right]^2\,.
\eenn
Figure~\ref{fig:snr_intgrand} shows the scales that dominate the Fisher integral $I(k)$. For comparison, the nonlinearity scale $k_{\rm NL}$, defined implicitly by $\int^{k_{\rm NL}}\dd k\,k^2 P_{\rm L}(k,z)/2\pi^2=1$, gives $k_{\rm NL}\simeq 10\,h\,\mathrm{Mpc}^{-1}$ at $z_{\rm eff}=5$. As seen from the figure, most of the signal originates from linear or weakly nonlinear scales $k\ll k_{\rm NL}$.

\clearpage
\bibliographystyle{apsrev4-1}
\bibliography{refs}

\end{document}